\documentclass[aps,prl,showpacs,amsmath,amsfonts,
               superscriptaddress,twocolumn]{revtex4}
\usepackage{graphicx}
\usepackage{bm}
\usepackage{pslatex}

\renewcommand{\eqref}[1]{Eq.(\ref{eq:#1})}

\begin{document}
\title{Loop updates for quantum Monte Carlo simulations in the canonical ensemble}

\author{S.M.A.~Rombouts, K.~Van~Houcke, L.~Pollet, }
\affiliation{Universiteit Gent - UGent, Vakgroep Subatomaire en Stralingsfysica
         \\
         Proeftuinstraat 86, B-9000 Gent, Belgium
         }
\date{\today}

\begin{abstract}
We present a new non-local updating scheme for quantum Monte Carlo simulations,
which conserves particle number and other symmetries.
It allows exact symmetry projection and  direct evaluation 
of the equal-time Green's function 
and other observables in the canonical ensemble.
The method is applied to bosonic atoms in optical lattices,
neutron pairs in atomic nuclei
and electron pairs in ultrasmall superconducting grains.
\end{abstract}

\maketitle


Quantum Monte Carlo methods (QMC) allow an in principle exact simulation of 
quantum many-body systems~\cite{Cepe99}.
Over the past decade, cluster updates have increased 
the efficiency of lattice-QMC methods dramatically~\cite{Evertz93,Evertz03}.
They proved particularly useful near phase transitions,
where the traditional algorithms suffered from a critical slowing 
down~\cite{Lind92}.
Based on an analogy with the Swendsen-Wang algorithm for classical 
systems~\cite{Swendsen87},
the first loop algorithm for QMC constructed clusters in the form of loops,
which then could be flipped to obtain new configurations~\cite{Evertz93}.
An important development was the formulation in continuous imaginary 
time~\cite{Beard96}, which eliminated discretization errors.
The worm algorithm~\cite{Prok98} established the link between the 
construction of the loops and the sampling in an extended configuration space,
thereby allowing a direct evaluation of the one-body Green's function.
The loop updates have also been implemented in the stochastic series expansion
method (SSE)~\cite{Sand99}.
A further optimization of the loop construction process was developed
in the form of {\em directed loops}~\cite{Sylj02}.
These turned out to be a special case of a more general {\em locally optimal} 
strategy, which tries to optimize the loop construction
by using the optimal stochastic transition kernel at each local step 
of the process~\cite{Poll04e}.
The forementioned loop algorithms all lead to fluctuations in particle number 
(or in magnetization for spin systems).
Therefore they sample the grand canonical ensemble.
There are cases where particle-number conservation plays an important role
and where one would like to sample directly the canonical ensemble:
e.g. lattice systems at commensurate fillings~\cite{Fisher89}
or finite systems such as ultrasmall superconducting grains~\cite{Duke99,Delf01}
or atomic nuclei~\cite{Koon97,Romb98}.
Results for the canonical ensemble can be obtained from loop algorithms by
by using only those configurations in the sample 
which have the right particle number~\cite{Roos99} 
or by explicitly throwing out the loop updates 
which change particle number~\cite{Evertz03}.
However, this is impractical because a lot of effort is wasted 
on the discarded configurations or updates
and because it requires a good estimate of the chemical potential.
In this letter, we present a class of loop updates 
which explicitly conserve particle number.
The algorithm results in moves that are always accepted,
which makes it easier to code and more efficient to run 
than other loop-update schemes.
Furthermore, one can impose the conservation of other symmetries.


Like most QMC methods, 
our algorithm starts from a decomposition of the imaginary time propagator,
$ U(\beta)=\exp(-\beta H)$. 
Generally one can write the Hamiltonian as $H= H_0 - V,$ 
consisting of an easy part $H_0$ and a residual interaction $V$
(note the minus sign in front of $V$, in order to ease notations further on).
For such a Hamiltonian, one can make a perturbative expansion in $V$ using
the following integral expression:
\begin{equation}
 U(\beta)  =  \sum_{m=0}^{\infty}  
 \begin{array}{l} \\ \int V(t_1) V(t_2) \cdots V(t_m) 
                                 e^{-\beta H_0} dt_1 dt_2 \cdots dt_m, \\
     {\scriptstyle 0 \leq t_1 \leq t_2 \leq \ldots \leq \beta}
 \end{array}
\end{equation}
with $V(t)=\exp(-t H_0) V \exp( t H_0)$.
Instead of sampling the representations of the operator $U(\beta)$ directly,
our method will perform a Markovian random walk 
in an extended configuration space, 
related to the decomposition  of the operator
$  U'(\beta,\tau) = e^{- \tau H} A e^{- (\beta -\tau) H}, $
where $A$ is the {\em worm operator}, to be defined later on.
An alternative would be to insert 
a creation and an anihilation operator at different imaginary times.
This forms the basis of the continuous-time loop algorithm \cite{Beard96}
and the worm algorithm \cite{Prok98}.
Here we will show that it is advantageous to work with a single worm operator,
provided that it commutes with the residual interaction: $AV=VA$.
If the operator $A$ furthermore commutes 
with the generator of a symmetry of $H_0$ and $V$,
then one can restrict the configurations to specific symmetry representations,
such that symmetry-projected results are obtained.
In particular one can sample the canonical ensemble
with a worm operator that conserves particle number.
 
By taking the trace (restricted to the wanted particle number and symmetry) 
and inserting complete sets of eigenstates of $H_0$ 
between all operators in the integral representation of $U'(\beta,\tau)$,
one assigns a weight $W(m,i,t,\tau)$ 
to the configurations specified by an order $m$, 
a set of inserted eigenstates $i_0, i_1, \ldots, i_m$, 
interaction times $t_1, t_2, \ldots, t_m$,
and the worm insertion time $\tau$. 
Let $i_L$ and $i_R$ denote the states to left and right of the worm operator.
We will call the configurations for which $i_L = i_R$  
{\em diagonal configurations}.
One can choose the worm operator $A$ such that its diagonal elements
are constant, i.e. $\langle i | A | i \rangle=c$ for all basis states $i$.
Then the sum of the weights of all the diagonal configurations is proportional 
to the particle-number projected trace of the operator $U(\beta)$,
which is nothing else than the {\em canonical partition function}.
Hence, the sampling of the configurations 
proportional to their weight $W(m,i,t,\tau)$ leads
to a sampling of the {\em canonical ensemble}.

To this end a Markov process is used 
for which the stationary distribution reflects
the contribution of the configurations to the extended partition sum
$Z'_N(\beta)=\mathrm{Tr}_{N}\left[U'(\beta,\tau)\right]$. 
Let us assume that a diagonal configuration is given.
We are free to choose a new insertion time $\tau$ for the worm operator 
because the weight is independent of $\tau$.
Then we perform a number of Markov steps according to the following rules,
until the move halts again in a diagonal configuration. 
We present the Markov rules in terms of a set of parameters
$q_D, c_D, {\mathcal N}_{DD'}, \epsilon_D, g_D, a_D$ and $s_D$, 
which are defined in detail in table \ref{tab:params}:
\begin{itemize}
\item 
      Evaluate the diagonal energy to the left $(E_L)$ 
      and to the right $(E_R)$ of the worm operator, 
      and evaluate the values of $q_L$ and $q_R$.
\item 
      Choose a direction $D$, either left (L) or right (R),
      proportional to the relative weights $q_L$ and $q_R$.
      Let $D'$ denote the opposite direction.
\item 
      With probability $c_D$, insert an interaction term $V$ 
      and choose a new intermediate state $i_i$ according to the distribution 
      $ P_{DD'}(i_i) = \langle i_D | A | i_i \rangle  
                          \langle i_i | V | i_{D'} \rangle
                      /  \langle i_D | A V | i_{D'} \rangle.$
\item 
      Move the worm in direction $D$ over a step size $\Delta \tau$ chosen
      according to a Poisson distribution: 
        $\Delta \tau \propto \exp(- \epsilon_D \Delta \tau )$.
      Assume periodic boundaries in imaginary time. 
\item 
      If the worm  would pass an interaction term in the time step 
      $\Delta \tau$,
      there are three options: 
      (a) With probability $s_D$: remove the interaction and halt the worm.
               Then the Markov step ends here. 
      (b) With probability $a_D$: remove the interaction, adjust the parameters
               and continue the worm move in the same direction.
      (c) With probability $1-a_D-s_D$, 
               or if the interaction can not be removed: 
               let the worm pass the interaction,
               but choose a new intermediate state according to $P_{DD'}(i_i)$.
               Adjust the parameters 
               and continue the worm move in the same direction.
\item 
       When the worm has reached the end of the time step $\Delta \tau$,
       then with probability $1-g_D$ the worm is halted 
       and the Markov step ends here. 
       With probability $g_D$, insert an interaction 
       and choose a new intermediate state
       according to $P_{DD'}(i_i)$.
       Adjust the parameters, draw a new time step $\Delta \tau$ 
       and continue the worm move in the same direction.
\end{itemize}
There is considerable freedom in the definition of the sampling parameters.
Guided by the principle of locally optimal moves~\cite{Poll04e,Houck05}, 
we propose the definitions as given in table~\ref{tab:params}.
These rules assure that the Markov chain satisfies 
the detailed balance condition
for the weight $R_{LR} W(m,i,t,\tau)$, with $R_{LR}=q_R+q_L$.
One could use the Metropolis-Hastings algorithm~\cite{Metr53,Hast70} 
in order to sample according to the weights $W(m,i,t,\tau)$.
Because the factors $R_{LR}$ fluctuate only mildly in practice,
one can accept all moves and take the extra weighting factor $R_{LR}$ 
into account when evaluating observables.
This speeds up the algorithm and reduces the complexity of the code.
Note that during a Markov step the worm keeps moving in the same direction,
even after passing, inserting or removing interactions.
This is possible while maintaining detailed balance 
because $A$ and $V$ commute.
Otherwise one would have to consider so-called {\em bounces}:
the worm could return on the path were it came from and undo its last changes.
This is one of the reasons why our new algorithm is more efficient
than the worm and loop algorithms 
of Refs.~\cite{Evertz93,Beard96,Prok98,Sylj02}.
%
\begin{table}[h]
\centering
\begin{ruledtabular}
  \begin{tabular} {|c||c|c|}
      parameters &  diagonal configurations & non-diagonal configurations \\
      ($ E_D \leq E_{D'} $) & $ (i_D =i_{D'}) $   & $ (i_D \neq i_{D'}) $ \\ 
   \hline
   \hline
    $   E_D   $ &  \multicolumn{2}{c|}{ $ \langle i_D | H_0 | i_D \rangle $} \\
    $  E_{D'} $ &  \multicolumn{2}{c|}
                                {$ \langle i_{D'} | H_0 | i_{D'} \rangle  $} \\
    $ {\mathcal N}_{DD'} $ &  \multicolumn{2}{c|}
               {$ \langle i_D | V A | i_{D'} \rangle 
                                      / \langle i_D | A | i_{D'} \rangle  $} \\
   \hline
    $  q_D    $ & $          \phi        $ & $             0             $  \\ 
    $  q_{D'} $ & $          \phi        $ & $       E_{D'}-E_{D}        $  \\
    $  c_D    $ & $            1         $ & $             0             $  \\
    $  c_{D'} $ & $            1         $ &    
            $ \min\left(1,\frac{{\mathcal N}_{DD'}}{E_{D'}-E_{D}}\right) $  \\
    $  \epsilon_D    $ & $            0         $ & $       E_{D'}-E_{D}  $ \\
    $  \epsilon_{D'} $ & $            0         $ &
                                             $             0             $  \\
    $  g_D    $ & $            0         $ & 
         $ \min\left(1, \frac{{\mathcal N}_{DD'}} {E_{D'}-E_{D}} \right) $  \\
    $  g_{D'} $ & $            0         $ & $             0             $  \\
    $  a_D    $ & $            0         $ & $             0             $  \\
    $  a_{D'} $ & $            0         $ & 
            $ \min\left(1,\frac{E_{D'}-E_{D}}{{\mathcal N}_{DD'}}\right) $  \\
    $  s_D    $ & $    \phi / {\mathcal N}_{DD'}   $ & 
         $  \min\left(1, \frac{E_{D'}-E_{D}}{{\mathcal N}_{DD'}} \right) $  \\
    $  s_{D'} $ & $    \phi / {\mathcal N}_{DD'}   $ & $      0          $  \\
    $  R_{DD'}$ & $         2 \phi       $ & $     | E_{D'}-E_{D} |      $  \\
  \end{tabular}
 \end{ruledtabular}
  \caption{Definitions for the sampling parameters, 
           assuming that $E_D \leq E_{D'}$
           (interchange $D$ and $D'$ otherwise).
	   The global parameter $\phi$ should be taken small 
           (such that $\phi \leq  {\mathcal N}_{DD'} $ 
           for all diagonal configurations)  but not zero,  
           to make sure that the worm halts in diagonal configurations 
           that are sufficiently decorrelated.
  \label{tab:params}}
\end{table}

While each Markov step is based on local changes, 
the chain of steps between two diagonal configurations 
corresponds to a global loop update: 
if one follows the creation and annihilation part of the worm operator 
as they move through the world-line representation of the configurations,
one sees that they describe a loop 
which closes again when the configuration becomes diagonal.
By keeping track of the worm operator at the intermediate steps,
one can collect statistics for the expectation values 
of non-diagonal operators,
similar to the way one evaluates the one-body Green's function 
in the worm algorithm~\cite{Prok98}.
The advantage of our method is that the operators 
are always evaluated at equal imaginary times,
leading to much better statistics for non-diagonal operators.
%

\begin{figure}[h]
\begin{center}
\includegraphics[height=8.6cm, angle=270]{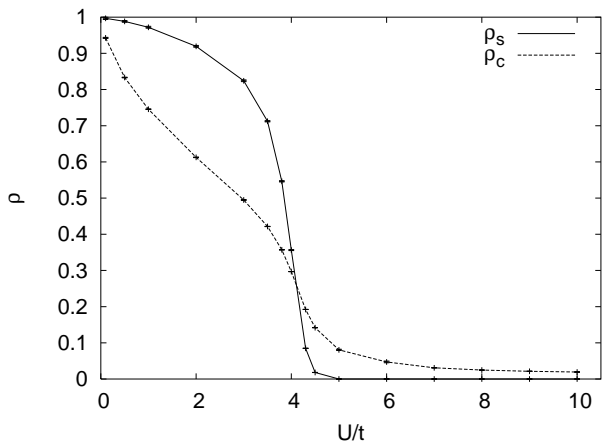}
\caption{Superfluid $(\rho_s)$ and condensed fraction $(\rho_c)$
         for the one-dimensional Bose-Hubbard model
         on a uniform lattice of 128 sites, 
         at an inverse temperature $\beta=128t^{-1}$.
         \label{fig:boshub}}
\end{center}
\end{figure}

As an illustration we have applied our method 
to the one-dimensional Bose-Hubbard Hamiltonian, given by
$  H= - t \sum_{i} b^\dag_i b_{i+1} + U \sum_i n_i(n_i-1)/2,$
with $n_i=b^\dag_i b_i$.
It models the low-energy degrees of freedom of various physical systems:
cold bosonic atoms in an optical lattice~\cite{Jaksch98},
$^4$He atoms on graphite~\cite{Zimanyi94}, 
and superconducting islands 
or grains connected by Josephson junctions~\cite{vanOudenaarden96}.
Various many-body techniques have been deployed to study its phase diagram.
Without trying to be complete, we mention here 
algebraic~\cite{Lieb63}, 
mean-field~\cite{Fisher89}, 
perturbative~\cite{Freericks94}, 
renormalization-group~\cite{Kuhn00} 
and QMC approaches~\cite{Scalettar91}. 
For our method we take 
$  V =  \sum_{i} b^\dag_i b_{i+1} $
and
$  A =  \sum_{i,j} b^\dag_i b_j $.
At commensurate fillings, this model can undergo a quantum phase transition
from a superfluid to a Mott isolator.
With grand-canonical methods it is difficult to simulate this transition
because one has to determine the chemical potential such 
that the exact density is obtained.
Fig.~\ref{fig:boshub} shows the superfluid 
and condensed fraction for a uniform one-dimensional
system of 128 sites at nearly zero temperature ($\beta=T^{-1}=128t^{-1}$) 
at a density of exactly one particle per site.
The superfluid fraction was derived from the winding numbers~\cite{Poll87},
while the condensed fraction was obtained 
from the one-body equal-time Green's function.

\begin{figure}[h]
\begin{center}
\includegraphics[height=8.6cm, angle=270]{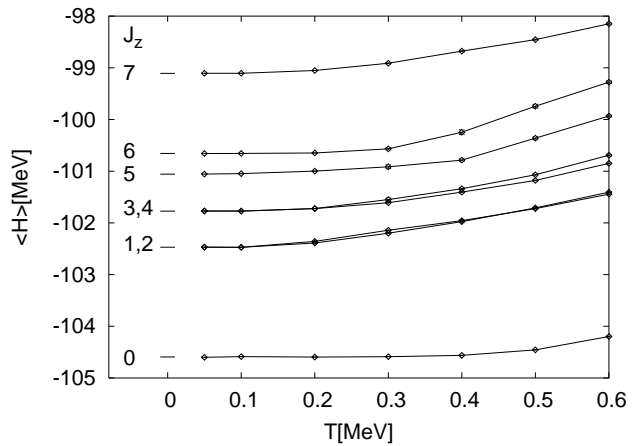}
\caption{ $J_z$-projected internal energies as a function of temperature
          for 10 neutrons in the $pf+sdg$ shell, 
          for a constant pairing interaction with $G=0.286$
          and single-particle energies as in Ref.~\cite{Romb98}.
	  On the left hand side the exact eigenvalues are shown~\cite{Romb04}.
        \label{fig:jzproj}}
\end{center}
\end{figure}

Fermionic models can be handled too, provided that there is a symmetry
which guarantees that all matrix elements are positive,
such that the sign problem is absent.
This is the case for the pairing Hamiltonian:
\begin{equation}
   H = \sum_{\alpha jm} e_j a^\dag_{\alpha jm}a_{\alpha jm} 
     - \frac{G}{4} \sum_{\alpha jm\alpha 'j'm'}  
             a^\dag_{\alpha jm} a^\dag_{\alpha j\bar{m}}
                 a_{\alpha 'j'\bar{m}'} a_{\alpha 'j'm'},
\end{equation}
where $a^\dag_{\alpha j m}$ creates a particle 
in a state with quantum numbers $\alpha, j, m$,
and $a^\dag_{\alpha j\bar{m}}$ a particle in the time-conjugated state.
The pairing model has been used extensively in nuclear physics
to model the pair correlations between nucleons and to explain
the particular odd-even effects 
found in nuclear spectroscopy~\cite{Heyde90,Dean03}.
Canonical QMC methods for the pairing Hamiltonian  
have been presented before:
Cerf's world-line QMC method~\cite{Cerf96} does not have a sign problem, 
but it does not sample the full phase space of broken pairs~\cite{Romb98c},
which makes it impractical for finite temperature calculations.
The shell model Monte Carlo method 
overcomes this problem~\cite{Koon97,Romb98,Romb99},
but has a sign problem for odd particle numbers.
For a constant pairing strength $G$, 
the eigenstates can be calculated exactly~\cite{Rich63,Romb04}.
Our method supplements these algebraic solutions 
with finite temperature results.
It can also be applied to level-dependent pairing interactions,
for which no algebraic solution is available. 
Here, we take $V$ to be the non-diagonal part of the pairing interaction
and we take $A$ equal to $V$ plus a constant diagonal term.
A pair-breaking term has to be added to $A$ and $V$
in order to ensure ergodicity~\cite{Houck05}.
Note that the worm operator conserves angular momentum.
Therefore one can restrict the intermediate states to a specific value 
of the quantum numbers $J$ and $J_z$.
Fig.~\ref{fig:jzproj} shows how the $J_z$-projected finite-temperature
results converge to the exact eigenvalues at low temperature,
for a model that describes neutron pair correlations 
in $^{56}$Fe~\cite{Romb98}.

\begin{figure}[h]
\begin{center}
\includegraphics[height=8.0cm, angle=270]{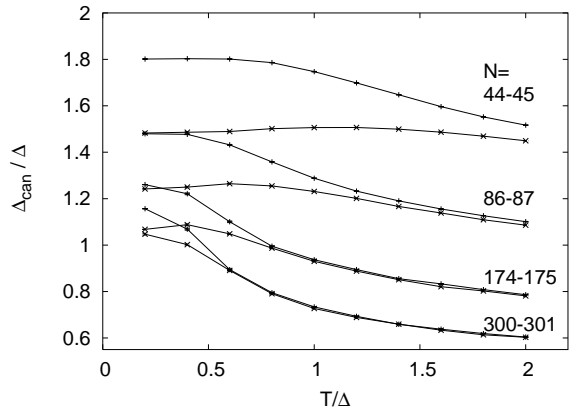}
\caption{ Canonical pairing gap $\Delta_{\mathrm{can}}$ 
       as a function of temperature,
       in units of the bulk gap $\Delta$,
       for the picket fence model at half filling 
       ($N$ particles), for various system sizes
       (44, 86, 174 and 300 levels) and 
       with pairing interaction strength $G=0.224d$ (as in Ref.~\cite{Sier00}).
        \label{fig:grains}}
\end{center}
\end{figure}

The pairing model where all levels are double degenerate 
($j=1/2$ for all $\alpha$) 
and equally spaced (with level spacing $d$), 
is called the {\em picket fence model}.
It applies to ultrasmall metallic grains, 
provided that a canonical approach is used~\cite{Duke99,Delf01,Sier00}.
One has found that grand-canonical approaches lead to an abrupt but unphysical
suppression of the superconductive correlations 
for level spacings larger than the bulk gap $\Delta$.
The extension of the exact ground-state calculations to finite temperatures
was cited as an open problem in Ref.~\cite{Delf01},
and has only been treated at the mean-field level~\cite{Ross01}
(except for the smallest systems).
Our method provides exact finite-temperature results 
both for odd and even particle numbers.
Therefore we can study the odd-even asymmetry,
which is a key indicator of superconductive correlations.
An illustrative quantity is the canonical pairing gap
$\Delta_{\mathrm{can}}$, as defined by Eq.(92) of Ref.~\cite{Delf01}.
It is a measure of the change in energy due to pairing correlations.
Its deviation from the BCS bulk gap indicates the
difference between the canonical and grand-canonical ensemble.
In Fig.~\ref{fig:grains} one sees the canonical pairing gap 
converging to the bulk gap with increasing system size.
There is a clear odd-even effect at lower temperatures
that disappears at higher temperatures,
and this transition temperature decreases with increasing system size.

These applications show the versatileness of the canonical loop updates.
Our method can be used to sample configurations with specific symmetries
and in particular to sample the canonical ensemble.
It leads to a very efficient sampling scheme with all moves accepted 
and without 'bounces' or critical slowing down.
It can be applied to bosons and fermion pairs
in discrete model spaces and to spin systems at fixed magnetization.
Off-diagonal observables such as the equal-time one-body Green's function 
can be evaluated with high efficiency.
This opens new perspectives for the study of quantum many-body systems
where particle number and other symmetries play an important role,
such as the nuclear shell model and ultrasmall superconducting grains.

The authors wish to thank K.~Heyde, J.~Dukelsky, S.~Wessel and M.~Troyer
for the interesting discussions
and the Fund for Scientific Research - Flanders (Belgium),
the Research Board of the University of Gent 
and N.A.T.O. for financial support.


\end{document}